\documentclass[pra,preprint,showpacs]{revtex4}

\def\mcl{\mskip-3mu}
\def\defeq{\mathop{\stackrel{\mathrm{def}}{=}}\nolimits}  
\usepackage{latexsym}
\usepackage[]{graphicx}
\usepackage{times}

\begin{document}

\title{Quantum-mechanical motion and the stillness of experimental records}

\author{F. Hadi Madjid}

\affiliation{82 Powers Road, Concord, Massachusetts 01742}   

\author{John M. Myers} 

\affiliation{Gordon McKay Laboratory, Division of Engineering and Applied
Sciences, Harvard University, Cambridge, Massachusetts 02138\\}

\date{19 May 2003}

\begin{abstract}
Experimenting with metastability in recording devices leads us to
wonder about an interface between equations of motion and the stillness
of experimental records.  Here we delineate an interface between wave
functions as language to describe motion and Turing tapes as language
to describe experimental records.  After extending quantum formalism to
make this interface explicit, we report on constraints and freedoms in
choosing quantum-mechanical equations to model experiments with
devices.  We prove that choosing equations of wave functions and
operators to achieve a fit between calculated probabilities and
experimental records requires reaching beyond both logic and the
experimental records.  Although informed by experience, a
``reach beyond'' can fairly be called a \textit{guess}.

Recognizing that particles as features of wave functions depend on
guesswork, we introduce their use not as objects of physical
investigation but as elements of thought in quantum-mechanical models
of devices.  We make informed guesses to offer a quantum-mechanical
model of a 1-bit recording device in a metastable condition.
Probabilities calculated from the model fit an experimental record of
an oscillation in a time-varying probability, showing a
temperature-independent role for Planck's constant in what heretofore
was viewed as a ``classical'' electronic device.

\end{abstract}

\maketitle

\section{Introduction}\label{sec:1}

Ubiquitous in physics is the need to decide which of two signals came
first.  Besides direct measurements, such as whether a particle has
been emitted prior to a clock tick, decisions about which of two
signals come first pervade computers.  In this connection we
experimented some years ago with a flip-flop as it confronted a race
between two signals.  As we see it now, the flip-flop is at once a
1-bit recording device and a 1-bit decision-making device.

In some essential uses, a flip-flop, after being written in, is read
twice.  It is known that exposure of a flip-flop to a race between a
signal and a clock tick sometimes results in a disagreement between
the two readings: one reading sees a 0 while the other sees a 1. The
frequency of disagreements is reported to decrease exponentially with
the waiting time $t$ between the race and the pair of readings
\cite{gray}; oscillation of the output voltage of a flip-flop has also
been reported \cite{chaney}, but to the best of our knowledge the
effect of this oscillation on disagreements has not been discussed. In
an experiment with flip-flops (Texas instruments 7400 and 74S00) we
found both the decay in frequency of disagreements and an oscillation
of that frequency.  Because of the oscillation, the decrease in
disagreement frequency is not monotonic: in some cases reading sooner
brings less chance of disagreement than reading later.

After more study of the recording of decisions about which of two
events happens first, we realized that this topic straddles an
interface between equations of motion and experimental records.  Here
we describe this interface and some of the consequences of recognizing
it.  Among these are a quantum-mechanical model of the flip-flop in
its dual role of deciding and recording, showing a role for Planck's
constant in computer circuitry, even at zero temperature.  

\subsection{A language interface}\label{subsec:1A}

Theoretical physics employs the mathematical language of continuous
functions to model motion, an example being a quantum-mechanical wave
function.  Experimental evidence, however, shows directly neither
motion nor continuous functions.  To be compared with theoretical
calculations, experiments must produce numbers, and these reside in
experimental records, about which one speaks in a different language
altogether.  The language of records, so far as theory is concerned,
is the language introduced by Turing, which chops motion into discrete
moves, interspersed by still moments in which a record---picture marks
on a Turing tape---can be read unambiguously, exactly because it is
devoid of motion.

By invoking probabilities, quantum mechanics makes room for both the
language of wave functions and that of records:  interpreted in terms
of quantum mechanics, experimental records express relative
frequencies, and the comparison between equations containing wave
functions and experimental records is a comparison between
probabilities calculated from the wave functions and experimental
numbers interpreted as relative frequencies. 

\subsection{Organization of the paper}\label{subsec:1B}

Section \ref{sec:2} displays the interface between wave-function language
and the language of Turing tapes that underlies descriptions of
experimental records.  As a category distinct from probabilities
calculated from wave functions and operators, experimental records
will be expressed mathematically, so as to display relative
frequencies of detections as these depend on how experimenters set
various knobs on instruments.  Then we extend quantum formalism to
make explicit the dependence of wave functions---or, more generally,
quantum states and operators---on parameters linkable by assumptions
to experimental knob settings.  As a consequence, probabilities
calculated from quantum states and operators depend on knob
parameters. In this way the interface between wave functions and
experimental records is made into an interface between a category of
functions of knob settings expressing probabilities calculated from
wave functions and a category of functions of knob settings expressing
experimental relative frequencies.  Putting these in separate
categories makes probabilities calculated from wave functions and
relative frequencies in experimental records show up as two distinct
legs on which quantum physics stands.

While both calculated probabilities and experimental relative
frequencies are functions of knob settings, there is unlimited variety
in both the probability functions of knob settings and the
experimental relative frequencies as functions of knob settings.  By
linking parameters of quantum-mechanical probabilities to knob
settings germane to a particular experiment, one makes what can be
called a quantum-mechanical model of the experimental setup.  In
this tying of theoretical probabilities to experiments, one can ask:
what constraint is imposed on a choice of models by requiring that the
model chosen generate probabilities that match relative frequencies
in a given experimental record?  As proved in Sec.\ \ref{sec:3}, a certain
freedom of choice endures no matter how tight a fit between model and
record is required.  While a scientist's choice of equations to model
an experiment is informed by logic and experimental data, the
scientist making this choice must reach outside of logic and data in
an act that amounts to a guess---albeit an informed guess.  
A wave function is a function of particle coordinates; as a corollary
to the proof, the particles expressed in a quantum-mechanical wave
function are open to choices unresolvable without guesswork.  Hence
the proof of freedom of choice in modeling establishes that quantum
physics stands not only on probabilities calculated from wave
functions and on experimental records but also on a third leg of
informed guesswork.

Although the format and content of experimental records requires for
its description not quantum language but the language of Turing tapes,
with their moments of stillness, one can ask about the dynamics of a
recording: what is the motion involved in putting a mark on a tape?
This is the subject of Secs.~\ref{sec:4} and \ref{sec:5}, in which we
navigate across the language interface.  By making an informed guess in
terms of particles we construct a quantum-mechanical model of a 1-bit
recording device and show its reasonably good agreement with an
experimental record.

\section{Formulating quantum mechanics for contact with
experimental records}\label{sec:2}

To compare an experiment with quantum-mechanical equations, one must
mathematically express features of things and acts in the laboratory.
We express these features only indirectly, as they are reflected in
experimental records, omitting from our consideration whatever
experimental activity escapes the record.

By an experimental \textit{record} we mean a digitized record, such as a
record stored in a computer, in contrast to an analog record.  By way
of explanation, the decision of which of two equations better fits an
experimental record requires extracting numbers from the record, as
does the unambiguous communication of a record from one scientist to
another.  If this extraction of numbers is postponed until after the
recording, the recording can be ``analog'', so that, for example, the
size of a number is read from the record by measuring the depth of an
impression in wax.  This makes the analog record like a foot print in
sand: if an experimental outcome is barely detectable, the record is
faint.  Sooner or later, however, the record has to be converted to
numbers, and it may as well be done in the making of the record.  A
recording device that does this is termed ``digital,'' and is
characterized electronically by regenerative amplification, so that a
signal that is too strong for a 0 and too weak for a 1 nonetheless
shows up, eventually, as one or the other.  That \textit{eventually} can
be like a coin landing on edge before it falls, and this is important,
but we ignore it until Sec.~\ref{sec:5}. To give mathematical expression
to experimental records, we think of records as binary coded, so that the
recording device is an array of unit devices, each of which contains a
single 0 or 1, like a marked square of a Turing tape \cite{turing}.

In some cases, an experimental record fails to match probabilities
calculated from one or another equation of quantum mechanics, and this
failure can indicate the need to choose a different equation.  To
clarify what is involved in choosing an equation to fit an
experimental record, a first step is to clarify what is supposed to
fit what, or, in other words, to clarify the interface between quantum
states and experimental records.

Locating this interface depends on recognizing that the mathematical
language of quantum mechanics stands in sharp contrast to ordinary
laboratory language.  When my colleague in the optics laboratory wants
to check the relative efficiency of detectors labeled A and B, he uses
ordinary language to ask me to swap A with B, and I do it.  This
enticing capacity of ordinary language to name particular things and
acts is entirely absent in mathematical language, which speaks only
about sets and functions as defined by relations among each
other. Because it talks only about itself, mathematics is indifferent
to its applications to physics.  We will take advantage of this
indifference when we use mathematical functions to describe
experimental records, a use categorically distinct from the use of
functions to express probabilities calculated from quantum states and
operators.  The purpose of this section is to display the interface
between calculated probabilities and experimental records as an
interface between functions used in these two categories, and to
extend quantum language to give this interface adequate recognition.
Displaying this interface by no means makes bridging it automatic.
Once the interface is in clear view, there is still the issue, to be
addressed in Sec.~\ref{sec:3}, of linking between a function expressing
experimental relative frequencies and a function expressing calculated
probabilities.

Without a theory to interpret it, no record can be read. For that
reason, to invoke mathematical functions to express records, we must
first touch on mathematical functions that express the states and
operators of quantum theory, along with the probabilities that ensue
from them.  Subsection \ref{subsec:2A} does this; following that, 
Subsec.\ \ref{subsec:2B} describes experimental records in relation to
knobs, \ref{subsec:2C} translates these knobs into mathematical functions
expressing states and operators, \ref{subsec:2D} describes classes of such
functions, and \ref{subsec:2E} shows how functions expressing quantum
states enter the design of experiments summarized by functions expressing
experimental records.

\subsection{Basic quantum language}\label{subsec:2A}
  
Our starting point to express probabilities calculable from quantum
states and operators is Dirac's formulation of quantum mechanics
\cite{dirac}, with the exception to be discussed in Sec.~\ref{sec:4} that
we see no essential need for state reductions.  Admittedly, this
language is narrow in the range of questions about experiments that it
allows one to ask, and one might hope for a broader formulation
in the future; however, for our purpose this narrowness of language is
an asset: once we show a logical looseness in choosing models even
within this narrow language, one can only expect even more looseness
in any future, more laboratory-oriented formulation of quantum
mechanics.  As formalized by von Neumann \cite{vN} and boosted by a
little measure theory \cite{rudin}, Dirac's language can be summarized
as follows.  Let $H$ be a separable Hilbert space, let $\rho$ be any
trace-class, self-adjoint operator of unit trace on $H$ (otherwise
known as a density operator), and let $M$ be a $\sigma$-algebra of
subsets of a set $\Omega$ of possible \textit{theoretical outcomes}.  By
\textit{theoretical outcomes} we mean numbers in a purely theoretical
setting, in contrast to experimental outcomes later to be
introduced.  Let $E$ be any projective resolution on $M$ of the
identity operator on $H$ (which implies that for any $\omega \in M$,
$E(\omega)$ is a self-adjoint projection \cite{rudin2}). Then
quantum-mechanical grammar assigns to $\rho$ and $E$ a probability
distribution $\mu$ on $M$ defined by
\begin{equation}\mu(\omega) = \mathrm{Tr}[\rho E(\omega)],
\label{eq:mu}
\end{equation}
where $\mu(\omega)$ is the probability of an outcome in the subset
$\omega$ of $\Omega$.  By \textit{probability} we mean a theoretical
number calculated from states and operators, a calculation that is
categorically distinct from any relative frequencies in experimental
records. Although not apparent in the abbreviated symbols in Eq.\
(\ref{eq:mu}), the numerical structures which $\rho$ and $E$
denote vary as one applies the form (1) to specific cases.

As the theoretical elements of quantum mechanics to connect most
directly to experimental records, probabilities---$\mu(\omega)$---are
the only available candidate: it is these rather than the density
operators or resolutions of the identities that can be compared
directly with relative frequencies in experimental records, shortly to
be discussed.

\subsection{Mathematical functions expressing experimental
records}\label{subsec:2B}

The functional form (\ref{eq:mu}) constrains the design of
experiments, or, more exactly, constrains quantum-mechanical
interpretations of experimental records.  Because quantum language is a
language of probability, its application to an experimental record
requires interpreting the record as a ``set of trials.''  In addition,
if an experimental record is to be compared with equations of the form
(\ref{eq:mu}), the record must be interpreted consistently with the
partition---sometimes called the ``Heisenberg cut''---between state
preparation, expressed by a wave function or density operator $\rho$,
and detection, expressed by a resolution of the identity $E$.  This
constraint of quantum grammar on the format of an experimental record
does not relegate a knob on a device once and for all to preparation
or detection, for the assignment to one or the other depends on the
experiment in which the device is employed and on how, within the
choices open in quantum mechanics, one chooses to interpret the
experimental record.

Regarding what is recordable from an experiment and relevant to
comparison with equations of quantum mechanics, we limit our attention
to names of actions taken and quantities associated with those actions
at each of a set of trials.  It is convenient to think of the
recordable actions as the setting of various knobs. Knobs can express
more than one might think.  For instance, because knobs can control
connectivity of components (e.g., by having some of the knobs
control a switching network) different knob settings can entail
different connections by wire, optical paths, etc., among the
components, and certain knob settings can effectively detach a
component from participating in an experiment.

It is convenient to call whatever numbers one learns from the
detectors once the knobs are set in a certain way for any one trial an
\textit{experimental outcome}.  (Note: by \textit{experimental outcome} we
mean numbers extracted from an experimental record; we do not mean an
eigenvalue or anything else calculated from states and operators.)
Then an experimental record conveys for each trial the names of knobs,
how each knob is set, names of detectors, and an experimental outcome
consisting of a numerical response for each detector.

We are thus led to express records mathematically in terms of
functions, the most primitive of which expresses the numerical outcome
of the detectors along with the knob settings for a given trial.  The
crucial point is that such functions express experimental
outcomes as a category distinct from theoretical probabilities.  From a
collection of such functions over a run of trials, one can compute for
any knob settings the relative frequencies of various
experimental outcomes.  These relative frequencies are our main focus.

To express relative frequencies extracted from an experimental record
so they can be interpreted quantum-mechanically, one partitions the
recorded knob settings into two sets, a set $A$ of knob settings that
one views as controlling state preparation and a set $B$ of knob
settings viewed as controlling detectors.  In addition, one stratifies
the possible experimental outcomes in a set $C$ of mutually exclusive
bins for tallying.  An experimental record, actual or hypothetical,
can then be summarized by a collection of relative-frequency
distributions, one for each setting of knobs $(a,b) \in A\times
B$. For the relative-frequency distribution on $C$ for knob settings
$(a,b)$, extracted from an experimental record $r$, we write
$\nu_r(a,b)$, so for any $c \in C$ we have $\nu_r(a,b)(c)$ for the
ratio of the number trials with knob setting $(a,b)$ and an
experimental outcome $c$ to the number of trials with knob setting
$(a,b)$ regardless of the outcome.  Thus one has
\begin{equation}(\forall a \in A, b \in B) \sum_{c\in C} \nu_r(a,b)(c) =
  1.
\label{eq:nudef}\end{equation}
\noindent\textit{Remarks}:
\begin{enumerate}
\item It is a matter of convenience whether to view $\nu_r$ as a
function-valued function with domain $A\times B$ or a function from $A
\times B \times C$ to the interval [0,1]. 
\item Without reference to any density operator or resolution of the
identity, one can express a question or an assertion about an
experimental record in terms of a knob-dependent relative frequency
$\nu_r$.  In this way the experimental record is expressed
mathematically in a category distinct from that of probabilities
calculated from states and operators.
\item
To allow for the possibility that some knob settings were never tried
in the experiment that generated a record, one can allow $\nu_r$ to be
a restriction of a function of the form (\ref{eq:nudef}), meaning it
is defined only on some subset of the domain $A \times B$.
\item An element of $A$ can be a list of knob settings, corresponding
to settings of individual knobs $A_1, A_2,\dots.$
\item Records---and relative frequencies $\nu$---can vary both
to reflect differences in experiments and, for a given experiment,
to reflect differences in the level of detail described.
\end{enumerate}

\subsection{Quantum language of ``knobs'' to allow comparison with 
experimental records}\label{subsec:2C}

Distinct from expressing an experimental record by a function $\nu$,
we now express by a function $\mu$ what might be called a
\textit{quantum-mechanical model} of the physics that generates such a
record. For this, let $H$, $M$, and $\Omega$ be as above. Introduce two
sets, mathematically arbitrary, denoted $A$ and $B$ (to be interpreted as
sets of knob settings). Let $\mathcal{D}$ be the set of functions from
$A$ to density operators acting on $H$.  (By density operators we mean
self-adjoint, trace-class operators with unit trace and non-negative
eigenvalues.)  Let $\mathcal{E}$ be the set of functions from $B$
to projective resolutions of the identity on $H$.  Then a
quantum-mechanical model is any pair of functions $(\rho,E)$, $\rho
\in \mathcal{D}$ and $E \in \mathcal{E}$.  We speak of any such pair
$(\rho_\alpha,E_\alpha)$ as a quantum-mechanical model $\alpha$.  To
such a model $\alpha$ there corresponds a function $\mu_\alpha: A
\times B \rightarrow $\{probability measures on $M$\}.  It is a matter
of convenience whether to view $\mu_\alpha$ as a probability-measure
valued function with domain $A\times B$ or a function from $A \times B
\times M$ to the interval [0,1].  When we view it the latter way, we
will express the probability of an event $\omega \in M$ with respect
to the measure corresponding to knob settings $(a,b)$ by
$\mu(a,b)(\omega)$.  The basic rule of quantum mechanics carries over
to the equation
\begin{equation}(\forall a \in A, b \in B, \omega \in M) \mu(a,b)
(\omega) = \mathrm{Tr}[\rho(a)E(b)(\omega)].
\label{eq:muab} \end{equation}

\subsection{Classes of quantum-mechanical models}\label{subsec:2D}

Within quantum mechanics one can set up various classes of
quantum-mechanical models of the form (\ref{eq:muab}).  For example, a
class can be characterized by the sets of knob-setting variables $A$
and $B$ and by an outcome space $\Omega$. Knobs allow definition and
analysis of some interesting relations among models of this class and
of some other classes, to be noted; knobs also allow definition and
analysis of some interesting relations among experimental records.

It is handy to name models so that, e.g., a model $\alpha$
involves a $\rho_\alpha$ along with an $E_\alpha$, with a consequent
$\mu_\alpha$.  If $\rho_\beta$ is a restriction of $\rho_\alpha$ (
meaning $A_\beta \subset A_\alpha$), and if $E_\beta$ is a restriction of
$E_\alpha$, we will say model $\beta$ is a restriction of model
$\alpha$ and that model $\alpha$ is an \textit{extension} of model
$\beta$.

As discussed in \cite{JOptB}, one can define morphisms between models
leading to the notion of a more detailed model \textit{enveloping} a less
detailed model.  The same notion can be applied to define how relative
frequencies visible in a more detailed report of an experiment envelop
those of a less detailed summary.  An informal example will be
given in Sec.~\ref{sec:4}.

It is also interesting and desirable to require less than an exact fit
between a model and an experimental record, which opens up questions
of distance between a model and an experimental record, as well as a
question of distance between one model and another.  For models
$\alpha$ and $\beta$ defined on the same domain $A \times B$, one can
define a distance between the two models in any of several ways, all
involving a notion of distance between two probability measures.  One
way is to use the maximum statistical distance between them
\begin{equation}
\max_{a \in A, b \in B} \mathrm{SD}[\mu_\alpha(a,b),\mu_\beta(a,b)],
\end{equation}
where $\mathrm{SD}$ is statistical distance \cite{statdist}.

Because models of the form (\ref{eq:muab}) are invariant under unitary
transformations, the essential feature of the states asserted by a
model $\alpha$ is the overlap between pairs of them.  The measure of
overlap convenient to Sec.\ \ref{sec:3} is
\begin{equation} \mathrm{Overlap}(\rho_\alpha(a_1),\rho_\alpha(a_2)) 
\defeq
\mathrm{Tr}[\rho_\alpha(a_1)^{1/2}\rho_\alpha(a_2)^{1/2}].
\label{eq:overlap}\end{equation}
One can measure the difference in state preparations asserted by
models $\alpha$ and $\beta$ for elements $a_1, a_2 \in A_\alpha \cap
A_\beta$ by $|\mathrm{Overlap}(\rho_\alpha(a_1),\rho_\alpha(a_2)) -
\mathrm{Overlap}(\rho_\beta(a_1),\rho_\beta(a_2))|$.

Note that an element $a \in A$ can be a list: $a=\{a^{(1)},\dots
,a^{(n)}\}$; in effect $A = A^{(1)} \times \cdots \times A^{(n)} =
A^{(1)} \times A'$, where $A' = A^{(2)} \times \cdots \times A^{(n)}$.
One can ``tape down a knob'' by setting it to a fixed value.  Suppose
for example that in a model $\alpha$ on $A \times B$ the knob setting
of $A^{(1)}$ is fixed at $a^{(1)}_0$. This fixing of a parameter
engenders a model $\beta$ on $A' \times B$; this model $\beta$ is a
restriction of model~$\alpha$.

In Sec.~\ref{sec:4} we will see models in which particles appear as
features of quantum states; two such models that have differing
numbers of particles can match the same summarized data.  This
fact will be shown to be important to a proper understanding of the
Heisenberg cut between state preparation and detection.

Any experimental record interpreted quantum-mechanically engenders a
class of models that fit that record.  Different models of this class
provide different theoretical pictures of the devices that take part
in the experiment, and the presence of a multiplicity of models may
call for extending the experiment in order to choose among them;
multiple models can also suggest different applications for the
devices modeled, and can prompt different modifications of the devices
for these applications.  An example will be discussed in Sec.\
\ref{sec:5}, in which two models share the knobs of an experimental
record but differ in their ``taped-down'' knobs.

\subsection{Dialog between functions used for theory
and functions expressing experimental records}\label{subsec:2E}
 
Not only do models supply language in which to make assertions about
the behavior of instruments, they are essential to the asking of
questions for experimental investigation.  Before one runs an
experiment one has to design it, which requires asking quantitative
questions.  To ask quantitative questions, one must: (a) have a
language that is numeric---i.e., mathematical, and (b) apply
that language to speak of things and acts on the laboratory bench,
such as the turning of knobs, not of themselves mathematical. In other
words, one uses models.  

Choosing one class of models rather than another opens up some avenues
of experimental investigation and bars others.  For example, choosing
of a class of models goes hand in hand with designing the format of
records for the experiment.  The record format is defined in terms of
experimental knob settings and bins for tallying experimental
outcomes.  In designing the experiment one organizes experimental knob
settings and bins for experimental outcomes in terms of what is
possible in terms of the chosen class of models \cite{ams}.  For
example, one links the set $C$ of possible patterns of detection
recorded for an experimental trial to the $\sigma$-algebra $M$ of
subsets of $\Omega$ for that class, so that to each $c \in C$ one
associates some $\omega_c \in M$.  In short, experimental outcomes can
be counted only in bins designed in terms of the theoretical outcomes
of some class of models.

Committing oneself to thinking about an experimental endeavor in terms
of a particular class of models makes it possible to:

\begin{enumerate}
\item borrow or invent models of that class;
\item organize experiments to generate data that can be compared with
models of the class;
\item express the results of an experiment without having to
assert that the results fit any particular model within the class;
\item pose the question of whether the experimental data fit one
model of the class better than they fit another model.
\end{enumerate}

\section{Choosing a model to fit a given experimental record}\label{sec:3}

Inverse to the problem of calculating probabilities from models
comprised of density operators and resolutions of the identity is
the problem of how given experimental records narrow our choice of
models. From an experimental record $r$ one can extract the
relative-frequency function $\nu_r$ (or a restriction of it) defined
on a set of possible knob settings $A \times B$.  Here we
explore: (1) constraints imposed on the choice of a quantum-mechanical
model $\alpha$ by requiring it to fit a given $\nu_r$, and (2) a
freedom of choice that survives the most stringent requirement
possible, namely an exact fit between a probability distribution
$\mu_\alpha$ and relative frequencies $\nu_r$ when these are defined
on a common domain $A \times B$.

For this limiting case of an exact fit, the experimental outcome set
$C$ generates the $\sigma$-algebra $M$ of subsets of $\Omega$, in the
sense that each $\omega \in \Omega$ is a disjoint union of elements
$\omega_c$ for $c \in C$.  Given empirical relative frequencies
$\nu_r: A\times B \times M \rightarrow [0,1]$, we ask what pairs of
functions $\rho_\alpha:A \rightarrow $ \{density operators on $H$\}
and $E_\alpha:B \rightarrow $\{projective resolutions of the
identity\} ``factor'' $\nu_r$ in the sense that
\begin{equation}(\forall a \in A, b \in B, \omega \in M)
  \nu_r(a,b)(\omega) = \mathrm{Tr}[\rho_\alpha(a)E_\alpha(b)(\omega)].
\label{eq:facnu} \end{equation}
Any such pair of functions $(\rho_\alpha,E_\alpha)$ constitutes a
quantum-mechanical model $\alpha$ with a probability function $\mu_\alpha$
that, in this extreme case, exactly matches the function $\nu_r$ that
expresses the experimental record.

\subsection{Constraint on density operators}\label{subsec:3A}

If for some values $a_1$, $a_2$, $b$ and $\omega$ one has
$\nu(a_1,b)(\omega)$ large and $\nu(a_2,b)(\omega)$ small, then
Eq.\ (\ref{eq:facnu}) implies that $\rho(a_1)$ is significantly
different from $\rho(a_2)$. This can be quantified in terms of the
\textit{overlap} of two density operators $\rho(a_1)$ and $\rho(a_2)$
defined in Eq.\ (\ref{eq:overlap}).  

\vspace{8pt}

\noindent\textbf{Proposition 1}: For a model $\alpha$ to be consistent
with relative frequencies $\nu$ in the sense of Eq.\ (\ref{eq:facnu}),
the overlap of density operators for distinct knob settings
has an upper bound given by
\begin{equation}
 (\forall a_1, a_2 \in A)\ \mathrm{Tr}[\rho(a_1)^{1/2}\rho(a_2)^{1/2}]
 \leq \min_{b,\omega} \{[\nu(a_2,b)(\omega)]^{1/2} +
 [1-\nu(a_1,b)(\omega)]^{1/2}\}.\label{eq:upper}
\end{equation} 

\vspace{8pt}

\noindent\textit{Proof}: For purposes of the proof abbreviate
$\rho(a_1)$ by $a_1$, $\rho(a_2)$ by $a_2$ and $E(b)(\omega)$ by $E$.
Because $E$ is a projection, $E = E^2$.  With this, the Schwarz
inequality (for any trace-class operators $F$ and $G$,
$|\mathrm{Tr}(FG^\dag)| \leq [\mathrm{Tr}(FF^\dag)]^{1/2}
[\mathrm{Tr}(GG^\dag)]^{1/2}$), and a little algebra one finds that
\begin{eqnarray} |\mathrm{Tr}(a_1^{1/2}a_2^{1/2})| &\mcl=\mcl&
|\mathrm{Tr}(a_1^{1/2}Ea_2^{1/2}) + \mathrm{Tr}(a_1^{1/2}(1-E)a_2^{1/2})|
 \nonumber \\ &\mcl \leq\mcl & |\mathrm{Tr}(a_1^{1/2}Ea_2^{1/2})| +
 |\mathrm{Tr}(a_1^{1/2}(1-E)a_2^{1/2})| \nonumber \\ &\mcl \leq \mcl&
 (\mathrm{Tr}\,a_1)^{1/2} [\mathrm{Tr}(Ea_2^{1/2}a_2^{1/2}E)]^{1/2} +
 [\mathrm{Tr}(a_1^{1/2}(1-E)(1-E)a_1^{1/2})]^{1/2} (\mathrm{Tr}\,a_2)^{1/2}
 \nonumber \\ &\mcl=\mcl & [\mathrm{Tr}(Ea_2^{1/2}a_2^{1/2}E)]^{1/2} +
 [\mathrm{Tr}(a_1^{1/2}(1-E)(1-E)a_1^{1/2})]^{1/2} \nonumber \\
&\mcl=\mcl&
 [\mathrm{Tr}(a_2E)]^{1/2} + [ 1 - \mathrm{Tr}(a_1E)]^{1/2}
 . \end{eqnarray} Expanding the notation, we have 
\begin{equation}
 (\forall a_1, a_2 \in A)\ \mathrm{Tr}[\rho(a_1)^{1/2}\rho(a_2)^{1/2}] \leq
 \min_{b,\omega}\{[\mathrm{Tr}(\rho(a_2)E(b)(\omega))]^{1/2} +
 [1-\mathrm{Tr}(\rho(a_1)E(b)(\omega))]^{1/2}\},\end{equation} which,
 with Eq.\ (\ref{eq:facnu}), completes the proof. $\Box$

\vspace{8pt}

\noindent\textit{Example}: For $0 \leq \epsilon,\delta \ll 1$, if for
some $b$ and $\omega$, $\nu(a_2,b)(\omega) = \epsilon$ and
$\nu(a_1,b)(\omega)= 1 - \delta$, then it follows that
$\mathrm{Tr}[\rho(a_1)^{1/2}\rho(a_2)^{1/2}] \leq \epsilon^{1/2} +
\delta^{1/2}$.  If, in addition, $\rho(a_1) = |a_1\rangle\langle a_1|$
and $\rho(a_2) = |a_2\rangle\langle a_2|$, then we have $|\langle
a_1|a_2\rangle|^2 \leq \epsilon^{1/2} + \delta^{1/2}$.

\subsection{Freedom of choice for density operators}\label{subsec:3B}

The preceding upper bound on overlap imposed by insisting that a model
agree with an experimental record invites the question: what positive
lower bound on overlap can be imposed by the same insistence?  The
answer turns out to be ``none,'' as we shall now state and prove.

While experimental records require only finite sets $A$ of knob
settings and $C$ of experimental outcomes for their expression, the
proposition to be stated holds also for these sets countably infinite
and with no restriction on the cardinality of the knob set $B$.  

\vspace{8pt}

\noindent\textbf{Proposition 2}: For the knob set $A$ and the set $C$
of experimental outcomes finite or countably infinite, and for any
$\nu:A\times B\times C
\rightarrow [0,1]$ satisfying Eq.\ (\ref{eq:nudef}), there exists a
Hilbert space $H$,
$\rho \in \mathcal{D}$ and $E \in \mathcal{E}$ such that

\vspace{8pt} 
1.\hfil${\displaystyle (\forall a \in A, b \in B, c \in C)
\nu(a,b)(c) = \mathrm{Tr}[\rho(a)E(b)(c)]}$,\hfil (10)\break
\addtocounter{equation}{1}

\vspace{-.75\baselineskip}
\vspace{8pt}
2.\hspace{.1in}$(\forall a)\ \rho(a)$ is a pure state $|a\rangle\langle
a|$ for some unit vector $|a\rangle$, and

\vspace{8pt}
3.\hspace{.1in}if $a \neq a'$, then
$\mathrm{Tr}[\rho(a)^{1/2}\rho(a')^{1/2}]
\equiv |\langle a|a'\rangle|^2 = 0$ so there is \textit{no} overlap.
\vspace{8pt}

\noindent\textit{Proof by construction}: Let $H$ be the direct sum of
countably infinite Hilbert spaces $H_a$, indexed by $a \in A$.  Let
$\rho(a) = |a\rangle\langle a|$ for some $|a\rangle \in H_a$, with the
consequence that
\begin{equation} (\forall a \neq a')\ 
\mathrm{Tr}[\rho(a)\rho(a')] \equiv |\langle a|a'\rangle |^2 = 0. 
\end{equation}
Let $E(b)(c) \defeq \sum_a \tilde{E}(a,b,c)$ with
$\tilde{E}(a,b,c)$ a self-adjoint projection on $H$ satisfying two
conditions:
\begin{enumerate}
\item $\tilde{E}(a,b,c)$ projects into  $H_a$, so that
\begin{equation}  \tilde{E}(a,b,c) |a'\rangle = 0\quad \text{if\ } a
\neq a' ,
\end{equation}
with the result that  
\begin{equation}
\mathrm{Tr}[\rho(a)E(b)(c)] = \mathrm{Tr}\left[\rho(a)
\sum_{a'}\tilde{E}(a',b,c)\right] = \mathrm{Tr}[\rho(a)
\tilde{E}(a,b,c)].
\end{equation}
\item For each pair $(a,b)$, let $\tilde{E}(a,b,c) =
|c(a,b)\rangle\langle c(a,b)|$ with unit vectors $|c(a,b)\rangle \in
H_a$  such that 
$\langle c(a,b)|c'(a,b)\rangle = \delta_{c,c'}$
and $|\langle c(a,b)|a \rangle |^2 = \nu(a,b)(c)$.
\end{enumerate}
This choice implies  $\nu(a,b)(c)
= \mathrm{Tr}[\rho(a)E(b)(c)]$, as promised. $\Box$

This proof shows the impossibility of establishing by experiment a
positive lower bound on state overlap without adding some
assumptions---or, as we say, guessing.  This has the following
interesting implication. Any experimental demonstration of quantum
superposition depends on showing that two different settings of the
$A$-knob produce states that have a positive overlap.  For example, a
superposition $|a_3\rangle = (|a_1\rangle +
e^{i\phi}|a_2\rangle)/\sqrt{2}$ has a positive overlap with state
$|a_1\rangle$.  Because, by Proposition 2, no positive overlap is
experimentally demonstrable without guesswork, we have the following:

\vspace{8pt}

\noindent\textbf{Corollary to Proposition 2}: Experimental
demonstration of the superposition of states requires resort to
guesswork.
\vspace{8pt}

\subsection{Constraint on resolutions of the identity}\label{subsec:3C}

Much the same story of constraint and freedom holds for
resolutions of the identity; in particular, we have:  

\noindent\textbf{Proposition 3}: In order for a model $\alpha$ to
fit relative frequencies $\nu_r$ in an experimental record,
the resolution of the identity must satisfy the constraint
$\| E_{\alpha}(b_1)(\omega_c) -
E_{\alpha}(b_2)(\omega_c) \| \ge \max_{a \in
A}|\nu_r(a,b_1)(c) - \nu_r(a,b_2)(c)|$.
\vspace{8pt}

\noindent\textit{Proof}: From the definition of the norm of an
operator $A$ as $\max_{\text{unit vectors } u}|Au|$, together with
showing that using density operators in place of unit vectors makes no
trouble, we have $\| E_{\alpha}(b_1)(\omega_c) -
E_{\alpha}(b_2)(\omega_c) \| \ge \max_{a \in
A}|\mathrm{Tr}[\rho_{\alpha}(a)E_{\alpha}(b_1)(\omega_c)] -
\mathrm{Tr}[\rho_{\alpha}(a)E_{\alpha}(b_2)(\omega_c)]| = \max_{a \in
A}|\nu_r(a,b_1)(c) - \nu_r(a,b_2)(c)|$. $\Box$

\subsection{Freedom of choice for resolutions of the
identity}\label{subsec:3D}

Can any positive upper bound less than 1 on
$\| E_\alpha(b_1)(\omega_c) - E_\alpha(b_2)(\omega_c)\|$ be
imposed by requiring that model $\alpha$ fit experimental relative
frequencies?  Even if the relative frequencies fit perfectly to a model
$\alpha$ for which $E_{\alpha}(b_1)(\omega_c)$ and
$E_{\alpha}(b_2)(\omega_c)$ have identical projections in $H_a$ for
all $a$ and all $c$, one can always map the (infinite-dimensional)
Hilbert space into itself by a proper injection, so that there is a
subspace $H_\perp$ orthogonal to all the $H_a$; then one can introduce
a model $\beta$ for which $E_{\beta}(b_1)(\omega_c)$ and
$E_{\beta}(b_2)(\omega_c)$ are as different as one wants in their
projections on $H_\perp$, so that for each $c$ one can have zero as
the projection of $E_{\beta}(b_1)(\omega_c)$ in $H_\perp$ and the unit
operator as the projection of $E_{\beta}(b_2)(\omega_c)$ in
$H_\perp$. Then $E_{\beta}(b_1)(\omega_c)$ and
$E_{\beta}(b_2)(\omega_c)$ differ by 1 in norm.  Thus we have:

\vspace{8pt}
\noindent\textbf{Proposition 4}: Measured relative frequencies can
impose no positive upper bound less than 1 on $\|
E(b_1)(\omega_c) - E(b_2)(\omega_c)\|$.

\vspace{8pt}

\noindent\textit{Remarks}:
\begin{enumerate}
\item More examples are given in \cite{ams} and \cite{JOptB}.
\item The proofs of Propositions 2 and 4 justify putting equations
involving states and operators in a category distinct from the
category of experimental records. On the state-and-operator side we
see motion expressed by a Schr\"odinger equation generating
probabilities; on the experimental side, we see ``tracks'' left in the
experimental record as static marks on a Turing tape.  These proofs
show that the two categories are incommensurable so that
guesswork is necessary to choose a model in terms of states and
operators as a theoretical expression of any experimental record.
\item 
In spite of the categorical difference between the quantum-mechanical
equations of motion and experimental records, we shall see in
Sec.~\ref{sec:5} that it is possible to use wave functions to model the
motion of writing a mark on a record.
\item If the sets $A$, $B$, and $C$ are finite,
then the argument goes through for $H_a$ and $H$ finite-dimensional.
\end{enumerate}

\section{Particles as elements of quantum language}\label{sec:4}

The language of quantum mechanics is a community endeavor: started by
Planck, shaped by Heisenberg and Schr\"odinger, infused with
probability by Born \cite{born}, made compatible with special
relativity by Dirac and later workers, and still developing. Having
recognized an interface between quantum states and operators as a
category and mathematical expressions of experimental records as a
distinct other category, we will offer in this section a
clarification of the grammar of quantum-mechanical language, to do
with the concept of a particle. 

Speaking of a ``particle'' one may be speaking about equations or
about an experiment, and a subtlety of physics language is that the
word \textit{particle} pertains to different things in these different
uses.  In this paper, by \textit{particle} we shall mean something
mathematical to do with a Schr\"odinger equation, or, potentially, its
generalizations.  In quantum-mechanical models particles appear both
in state preparation and in detection.  In both cases they appear as:
\begin{enumerate}
\item mass parameters,
\item position coordinates in configuration space, and
\item momentum coordinates in momentum space.
\end{enumerate}
For example, in nonrelativistic quantum mechanics, one sees mass
parameters and coordinates in any wave function, and hence in a joint
probability of detection; e.g., $n$ particles entail a
$3n$-dimensional volume element
$d^3\mathbf{x}^{(1)}\dots d^3\mathbf{x}^{(n)}$ where $d^3\mathbf{x}^{(j)}
= dx_1^{(j)}dx_2^{(j)}dx_3^{(j)}$, along with a joint probability
density $|\psi(\mathbf{x}^{(1)},\dots,\mathbf{x}^{(n)},t)|^2$ for
detecting these $n$ particles at time $t$.

To see the applicability of the preceding proofs to wave functions,
recall that a wave function is an element of a linear function space
and hence a state vector, to which the proofs apply: in order to
connect wave functions to experiments it is necessary to make guesses.
Recognizing this necessity has an impact on how we think of
``particles.''  Because particles appear in models only as features of
wave functions and of resolutions of the identity (and their
relativistic generalizations), it follows from Propositions 2 and 4
that particles can become associated with an experimental record only
by resort to guesswork.
  
The familiar particles discussed in field theories were invented in
connection with collision experiments, typically made with
accelerators.  Reports of collision experiments emphasize features
of experimental records that support or refute theories of particles
while downplaying the dependence of these features on the devices
and their knobs that constitute the accelerator.  We now propose
to put particles to work as terms in language for framing hypotheses
about devices controlled by various knobs, including devices having
little or nothing to do with collision experiments.  From the
propositions of Sec.\ \ref{sec:3} we conclude the following:

\vspace{8pt}

\noindent\textbf{Corollary}: As features of mathematical models,
particles to describe a device cannot be experimentally determined
without resort to guesswork.

\vspace{8pt}

In particular, one may need to change the particles by which one
models a device with changes in the context in which the device is
used. Recognizing it as a feature of models has an impact on the concept
of a particle, to which we now turn.

\subsection{No detecting the same particle twice}\label{subsec:4A}

Having recognized particles as features of quantum-mechanical language
used in conjunction with guesswork to describe experiments with
devices, we want to examine how these particles relate to the
structure of this language---its \textit{grammar}, one might say.  The
central point is that particles are constructs introduced to account for
experimental records of detections.  Changes in the experimental knob
settings can correspond to a change in parameters in a
quantum-mechanical model, and hence can correspond to a change in the
particles that are features of that model.  But what about detections?
Can an experimental detection correspond to a change in
particles as features of a quantum-mechanical model?  We shall propose a
rule of grammar that keeps particles of a given model in a category
isolated from that of experimental detections, making it a mistake in
grammar to suppose that detection can act like a change in a knob setting
that changes a wave function or a particle.

The rule to be proposed draws on an understanding not only of particles
as they enter a single model, but also of how particles are expressed in
multiple models having a variety of levels of detail.  Whenever a model
$\alpha$ entails the detection of a particle, say one with position
coordinates
$\mathbf{x}$, there can be a more detailed model $\beta$ that introduces
more particles
$\mathbf{y}_1$,
$\mathbf{y}_2, \dots,$ that act as probes of particle $\mathbf{x}$, so
that what model $\alpha$ expresses as the detection of $\mathbf{x}$ is
expressed by model $\beta$ as detections of probes $\mathbf{y}_1$,
$\mathbf{y}_2, \dots,$ in place of or in addition to $\mathbf{x}$.  With
this in mind, to speak of the effect of detection ``on a particle
$\mathbf{x}$'' of model $\alpha$ is either to speak nonsense, or to allude
to the related but different model $\beta$ in which at least one
additional particle $\mathbf{y}$ enters as a probe of $\mathbf{x}$.  If
this is done, one can speak within model $\beta$ of a (time-dependent)
joint probability density for theoretical outcomes for $\mathbf{y}$ and
$\mathbf{x}$.  Such a joint probability of course engenders a conditional
probability of detecting particle $\mathbf{x}$ given a theoretical
outcome for particle $\mathbf{y}$.  If $\mathbf{y}$ is viewed as probe
of $\mathbf{x}$, one can then say that the probe outcome affects the
probability distribution for $\mathbf{x}$, but this is no causal
relation, only the working of the mathematical rules for joint and
conditional probabilities.

In particular it is a mistake in the language rules of
probability to suppose that finding an outcome for particle
$\mathbf{x}$ changes the probability distribution for $\mathbf{x}$.
If one recognizes this, and if one sees wave functions as constructs
of a probability calculus, there can be no `change in the wave function
for a particle brought about by detecting that particle.'  

Recognizing that ``multiple detections of a particle'' is shorthand for
using probes, we propose as a rule of the grammar of quantum
language:

\vspace{8pt}

\noindent\textbf{Rule}: Within a given model, there can be no
possibility of more than one outcome per particle.

\vspace{8pt}

By way of touching base with some history, once one has on hand
conditional probabilities for theoretical outcomes for
$\mathbf{x}$ given theoretical outcomes for $\mathbf{y}$, it may be
tempting to cook up a wave function that can be used in an equation of
the form (\ref{eq:mu}) or (\ref{eq:muab}) to express that
conditional probability.  While there is nothing mathematically
incorrect in this maneuver, it puts the mind on a slippery slope to
supposing that ``detection of a particle'' changes not only the
probabilities of detecting other particles, but also the probability
of a second detection of the same particle; i.e. that detection makes
a change (a.k.a.\ ``reduction'') of the wave function.  To us, this
thought is now best viewed as a confusion stemming from inattention to
the grammar of quantum-mechanical language.\looseness=-1

\section{Application: a particle model of a 1-bit 
recording device}\label{sec:5}

Some experiments, instead of employing recording devices only to hold
a record for future examination, employ recording devices that are
read immediately, say as part of a feedback loop \cite{hj93}.
Describing the reading of a recording device immediately as it is
written into calls for special skill in navigating the linguistic
interface between experimental records and equations of quantum
dynamics.  We say \textit{linguistic} because this interface sits between
the language of Turing tapes and that of wave functions.  On one side
we have language for describing stillness, and on the other side
language for describing motion.  As will be shown, it would be
misleading to use Turing language to speak of a recording element in
this situation of prompt reading.  In other words, while Turing
language is appropriate for records in many contexts, it cannot be a
universal ``language of recording.''

To model a situation of reading soon after or in the midst of
recording, one needs to construct a model involving the motion of a
continuous field, not just the `moves' allowed by Turing language.
Correspondingly, to experiment with a recording device, one needs to
separate the recording device that is the subject of the experiment,
call it $F0$, from other recording devices, say $F1$, $F2$, etc.,
that are used to hold the experimental record of the investigation of
$F0$.

In this section we model the motion involved in writing and promptly
reading a record, and compare the model with the experimental record
obtained some years ago.  In this experiment, we investigated what
happens to a 1-bit recording device made of transistors---a
flip-flop---when it is exposed to a race between an arriving signal
and a clock tick announcing a deadline.  

We digress to remark that there is no sense in speaking of a lens as
classical or quantum, for the same glass lens can be used in an
experiment regardless of whether one chooses quantum or classical
language to design or report on the experiment, and the same holds for
a flip-flop constructed of transistors.  Although our experiment was
designed using classical electronics language, there is nothing to
stop our re-examining the experimental record in terms of
quantum-mechanical language.  When we do this, we will find new
questions and new directions for further experimental and theoretical
investigation.

\subsection{Metastability in the motion of records}\label{subsec:5A}

A 1-bit recording device can be asked to decide which comes first to
it, a signal to be recorded or a clock tick announcing a deadline.  For
this use, the 1-bit recording device is connected to whatever
generates the signal via a logic gate that acts as a switch by
breaking the signal path when a clock ticks (electronically) to
announce the deadline.  This situation of a race between a signal and
a clock tick is unavoidable in the design of computer-based
communications, and has long been of interest to hardware designers.
The breaking of the signal path by the clock can happen just as the
signal is arriving, reducing but not eliminating the surviving signal
that actually reaches the 1-bit recording device. This so-called
``runt pulse'' can put the recording device into what is termed a
metastable condition, like a tossed coin that happens to land on edge
and lingers, perhaps for a long time, before it shows an unambiguous
head or tail.  This possibility of a lingering metastable condition
threatens more than arbitrariness in the decision to record a 1 or a
0; the critical issue is that two readings of a metastable record can
be in conflict with one another: one can read a 0 while the other reads a
1, in effect splintering the abstraction that describes the contents
of a recording device as a number.

\subsection{Experimental design}\label{subsec:5B}

We investigated a 1-bit recording device implemented in
transistor-transistor logic circuitry as the flip-flop $F0$ shown in
Fig.\ \ref{fig:1}.  In addition to power and ground leads not shown, the
flip-flop has wire leads for input, clock, and read-out; it also has a
lead that works as a reset knob when driven by a low voltage.  The
NAND gate into which both the clock and the signal feed acts as a
switch.  At each trial $F0$ is exposed to a race between a signal
rising from low to high on the input line and the clock falling from
high to low, which turns off the switch just as the signal is passing
through it, resulting in a negative runt pulse emerging from the
switch.

Besides the flip-flop $F0$ subjected to a race, two other flip-flops,
$F1$ and $F2$ shown in Fig.\ \ref{fig:2}, are used to record the result of
each trial. The key experimental knob controls the delay time $t$ between
exposing $F0$ to the race and a pulse on the clock-1 line that causes the
reading of
$F0$ by the paired flip-flops $F1$ and $F2$.  After they read $F0$ and
after an additional delay to assure that they have settled down, the bits
recorded by $F1$ and $F2$ are tallied for later perusal.  Figure
\ref{fig:3} shows the experimental record of relative frequency of
disagreement between $F1$ and $F2$ as a function of the waiting time $t$
between exposing $F0$ to the race and its reading by $F1$ and $F2$. 
While the flip-flops $F1$ and $F2$ and the other registers to which these
transfer their results can be discussed in Turing language, the record
displayed in Fig.~\ref{fig:3} shows this is not the case for $F0$ exposed
to a race condition.

\subsection{Quantum-mechanical model of a metastable recording
device}\label{subsec:5C}

To model the metastable flip-flop $F0$ quantum-mechanically, notice that
although recording devices are ordinarily assigned in models to the
`detection' side of the Heisenberg cut, there is no necessity for this,
and in the present case, where $F0$ is part of what is being
investigated, we can and do choose to model it as belonging to state
preparation.  In contrast, the flip-flops $F1$ and $F2$ that read $F0$
continue to be assigned to the detection side.

We invoke the rule of Sec.~\ref{sec:4}, according to which the modeling of
whatever involves several detections must involve several particles.
The flip-flop $F0$ is to be written into, and, before any later rewriting,
it is to be read some number of times, say $n$ times.  We model each
reading as a detection, and hence are obliged by our rule to model the
writing in the device as preparing at least $n$ particles, so that
there can be $n$ detections.

Multiple readings entail the possibility of disagreements among them.
Within the computer language of Turing tapes, any single reading
can be expressed by a binary-valued variable $r \in \{0,1\}$.
Allowing the possibility of disagreements, the expression of $n$
readings of 1 bit is then a bit string $\vec{r} = (r_1,\dots, r_n)
\in \{0,1\}^n$.  To connect this Turing language to the laboratory
record, one has to partition the possible experimental outcomes into
(at least) $2^n$ blocks and interpret these as corresponding one-to-one to
the $2^n$ possible values of $\vec{r}$.  
 
To model a 1-bit recording device, we focus not on experimental
outcomes, but on outcomes in the mathematical sense of subsets of
$\Omega$ to which probabilities are assigned by an equation of the
form (\ref{eq:muab}).  Thus in our model $\alpha$ we will associate to
each $\vec{r}$ a subset $\omega(\vec{r}) \in \Omega$ with the notion
that $\mu_\alpha(a,b)(\omega(\vec{r}))$ is the probability asserted by
the model $\alpha$ for the outcome $\vec{r}$, given knob settings $a$
and $b$. With this in mind we propose a class of quantum-mechanical
models of a 1-bit recording device, guided by the following
considerations:
\begin{enumerate} 
\item We model the recording of a bit---\textit{writing}---as the
preparation of a wave function satisfying a Schr\"odinger equation.
\item Writing of a bit that can be read some number $n$ times
 is modeled by the preparation of a $n$-particle
 wave function.
\item Each of the $2^n$ possible results of $n$ readings of 1 bit
corresponds to one of $2^n$ mutually disjoint subsets of 
(the mathematical) outcome space $\Omega$.  
Of special interest are the two subsets $R_0,R_1 \subset \Omega$,
interpreted as follows: the probability $\mu(R_0)$ is taken
to be the probability of an $n$-fold reading $r_\ell = 0, \ell = 1,
\dots, n$, and similarly the probability $\mu(R_1)$ is taken
to be the probability of an $n$-fold reading $r_\ell = 1, \ell = 1,
\dots, n$.
 \begin{enumerate}
 \item The (unambiguous) writing of a 0 is modeled by the preparation
   of a $n$-particle wave function $\psi_0$ which evolves over time to
   produce a probability near 1 for an outcome in
   region $R_0$.
 \item The writing of a 1 is
   modeled by the preparation of a different wave function $\psi_1$
   which evolves to produce a probability near 1 for detecting $n$
   particles in region $R_1$.
 \end{enumerate}
\item We model a race condition as follows:
\begin{enumerate}
\item Writing prepares an $n$-particle wave function intermediate
between that for a 0 and that for a 1.
\item Discrepancies of readings of the 1-bit device correspond
to detections of $n$ particles in regions other than $R_1$ and $R_0$.
\item An interaction term in the Schr\"odinger equation for the
$n$-particle wave function couples the particles to produce a growth
over time in correlation in the $n$-particle detection probability,
meaning that the outcome regions other than $R_0$ and $R_1$ become
progressively less likely.
\end{enumerate}
\end{enumerate}

\vspace{8pt}

\noindent\textit{Remark}: The experimental records of the race
condition do not accord with any model we know in which the wave
function for the race condition is a quantum superposition of $\psi_0$
and $\psi_1$; thus the $\psi_{\mathrm{race}}$ is intermediate between
$\psi_0$ and $\psi_1$ but is not a superposition of them.

\vspace{8pt}

For simplicity, we consider only a single space dimension for each of
$n= 2$ particles. Let $x$ be the space coordinate for one particle and
$y$ be the space coordinate for the other particle (so $x$ and $y$ are
coordinates for two particles moving in the same single space
dimension, not for one particle with two space dimensions).  The
region $R_0$ is the region $x,y < 0$, the region $R_1$ is the region
$x,y >0$.  Notice that this leaves two quadrants of the $(x,y)$-plane
for discrepancies in which one reading yields a 0 while the other
reading yields a 1. The model proposed will be analyzed to see what it
says about the probability of discrepancies between paired detections,
and how this probability varies over time.

We now specify the 2-particle Schr\"odinger equation which is the
heart of this model.  Characterizing a 1-bit recording device by an
energy hump and assuming that what matters about this hump for
long-time settling behavior is its curvature, we start by expressing
uncoupled particles traveling in the presence of a parabolic energy
hump.  A single such particle, say the $x$-particle, is expressed by
\begin{equation}i\hbar\,\frac{\partial \;}{\partial t}\,\psi(x,t) =
\left(-\frac{\hbar^2}{2m}\,\frac{\partial^2 \;}{\partial x^2} -
\frac{k x^2}{2}\right)\psi(x,t).
\label{eq:single}\end{equation}  Equation (\ref{eq:single}) is the 
quantum-mechanical equation for an unstable oscillator, the
instability coming from the minus sign in the term proportional to
$x^2$.  To model a 1-bit recording device that can be read twice we
augment this equation by adding a $y$-particle. The effect of the
energy hump on this particle is then expressed by a term $-ky^2/2$ in
the hamiltonian.  In order to produce growth over time in the correlation
of the detection probabilities, we put in the coupling term
$\frac{1}{4}\lambda(x-y)^2$.  This makes the following two-particle
Schr\"odinger equation:
\begin{equation}i\hbar\,\frac{\partial \;}{\partial t}\,\psi(x,y,t) = 
-\frac{\hbar^2}{2m} \left(\frac{\partial^2 \;}{\partial x^2}
+\frac{\partial^2 \;}{\partial y^2}\right)\psi(x,y,t) +
\frac{k}{2}\left(-x^2 - y^2 + \frac{\lambda}{2}
(x-y)^2\right)\psi(x,y,t). \label{eq:pairphys}
\end{equation} 
The natural time parameter for this equation is $\omega^{-1}$ defined
here by $\omega \defeq \sqrt{k/m}$; similarly there is
a natural distance parameter $\sqrt{\hbar/m\omega}$.  

\subsection{Initial conditions}\label{subsec:5D}

For the initial condition, we will explore a wave packet of the form:
\begin{equation}
\psi(x,y,0) = \frac{1}{\pi^{1/2}b}\exp[-(x-c)^2/2b^2] \exp[(y-c)^2/2b^2].
\label{eq:initxy}
\end{equation}
{}For $c=0$, this puts the recording device exactly on edge, while
positive or negative values of $c$ bias the recording device toward 1 or
0, respectively.

\subsection{Solution}\label{subsec:5E}

As discussed in Appendix A, the solution to this model is 
\begin{equation}
|\psi(x,y,t)|^2 = \frac{1}{\pi B_1(t)B_2(t)}
\,\exp\left\{-\left(\frac{x+y}{\sqrt{2}} - c\sqrt{2}\cosh
  t\right)^2\Bigg/B^2(t)- \frac{(x-y)^2}{2B^2_2(t)}\right\},
\label{eq:jointxyphys}
\end{equation}
with
\begin{eqnarray}
B_1^2(t) &\mcl=\mcl& b^2 \left[1 + \left(\frac{\hbar^2}{\omega^2 m^2 b^4}
+1\right)\sinh^2 t\right]
\nonumber \\
B_2^2(t) &\mcl=\mcl& b^2\left[1 + \left(\frac{\hbar^2}{\omega^2 m^2
b^4(\lambda-1)}-1\right)
  \sin^2 \sqrt{\lambda-1}\,t\right].
\label{eq:Bphysdef}
\end{eqnarray}

The probability of two detections disagreeing is the integral of this
density over the second and fourth quadrants of the $(x,y)$-plane.
For the especially interesting case of $c =0$, this integral can be
evaluated explicitly as shown in Appendix A:
\begin{equation}
\Pr(F1\text{ and }F2\text{ disagree at }t) =
\frac{2}{\pi}\tan^{-1}\left(
\frac{\displaystyle 1 + \left[\frac{\hbar^2}{\omega^2 m^2
b^4(\lambda-1)}-1\right]\sin^2\sqrt{\lambda-1}\,\omega t}{\displaystyle 1
+\left(\frac{\hbar^2}{\omega^2 m^2 b^4}+1\right)\sinh^2 \omega
t}\right)^{1/2}.
\label{eq:edge1phys}
\end{equation}
This formula works for all real $\lambda$. For $\lambda > 1$, it shows
an oscillation, as illustrated in Fig.~\ref{fig:3}.  For the case $0 <
\lambda <1$, the numerator takes on the same form as the denominator, but
with a slower growth with time and lacking the oscillation, so that the
probability of disagreement still decreases with time, but more
slowly.  Picking values of $b$ and $\lambda$ to fit the experimental
record, we get the theoretical curve of Fig.~\ref{fig:3}, shown in
comparison with the relative frequencies (dashed curve) taken from the
experimental record.  For the curve shown, $\,\lambda = 1.81$ and $b =
0.556$ times the characteristic distance $\sqrt{\hbar/\omega m}$.
According to this model $\alpha$, a design to decrease the half-life
of disagreement calls for making both $k/m$ and $\lambda$ large.
Raising $\lambda$ above 1 has the consequence of the oscillation,
which can be stronger than that shown in Fig.~\ref{fig:3}.  When the
oscillation is pronounced, the probability of disagreement, while
decreasing with the waiting time $t$, is not monotonic, so in some 
cases judging sooner has less risk of disagreement than judging later.

\subsection{An alternative to model $\alpha$}\label{sec:5F}

Proposition 2 of Sec.~\ref{sec:3} asserts that whenever one model
works, so do other quite different models, and indeed we can construct
alternatives to the above model $\alpha$ of a 1-bit recording device.
Where model $\alpha$ distinguishes an initial condition for writing a
0 from that for writing a 1 by ``placing a blob,'' expressed in the
choice of the value of $c$ in Eq.\ (\ref{eq:initxy}), a model $\beta$
can distinguish writing a 0 from writing a 1 by ``shooting a particle
at an energy hump'' with less or more energy of wave functions
initially concentrated in a region in which $x$ and $y$ are negative
and propagating toward the energy saddle at $x,y = 0$.  Model $\beta$
can preserve the interpretation of detection probabilities (e.g.,
$\,x,y > 0$ corresponds to both detectors reading a 1).  Hints for
this construction can be found in the paper of Barton \cite{barton},
which contains a careful discussion of the energy eigenfunctions for
the single inverted oscillator of Eq.\ (\ref{eq:phiprob}), as well as
of wave packets constructed from these eigenfunctions.

Such a model $\beta$ based on an energy distinction emphasizes the
role of a 1-bit recording device as a decision device: it ``decides''
whether a signal is above or below the energy threshold.  For this
reason, energy-based models of a 1-bit recording device exposed to a
race condition can be applied to detectors working at the threshold of
detectability.  It would be interesting to look for the behavior shown
here for recording devices, including the oscillation, in a
photodetector used to detect light at energies at or below a
single-photon level.

\subsection{The dependence of probability of disagreement on
$\hbar$}\label{subsec:5G}

For finite $b$, the limit of Eq.\ (\ref{eq:edge1phys}) as $\hbar
\rightarrow 0$ is
\begin{equation}
\Pr(F1\text{ and }F2\text{ disagree at }t) =
\frac{2}{\pi}\tan^{-1}\left(\left|
\frac{\cos \sqrt{\lambda-1}\,\omega t}{\cosh \omega t}\right|\right).
\label{eq:edgeh0}
\end{equation}
This classical limit of model $\alpha$ contrasts with the
quantum-mechanical Eq.\ (\ref{eq:edge1phys}) in how the disagreement
probability depends on $\lambda$.  Quantum behavior is also evident in
entanglement exhibited by the quantum-mechanical model.  At $t = 0$ the
wave function is the unentangled product state of Eq.\
(\ref{eq:initxy}).  Although it remains in a product state when viewed in
$(u,v)$-coordinates discussed in Appendix A, as a function of
$(x,y)$-coordinates it becomes more and more entangled with time, as
it must to exhibit correlations in detection probabilities for the
$x$- and $y$-particles.  By virtue of a time-growing entanglement and
the stark contrast between Eq.\ (\ref{eq:edge1phys}) and its
classical limit, the behavior of the 1-bit recording device exhibits
quantum-mechanical effects significantly different from any classical
description.

The alternative model $\beta$ based on energy differences can be
expected to depend on a \textit{sojourn time} with its interesting
dependence on Planck's constant, as discussed by Barton \cite{barton}.
Another variation in models, call it $\gamma$, would distinguish
between 0 and 1 in terms of momenta rather than location; it too is
expected to have a robust dependence on $\hbar$.  These models thus bring
Planck's constant into the description of decision and recording
devices, not by building up the devices atom by atom, so to speak, but
by tying quantum mechanics directly to the experimentally recorded
relative frequencies of outcomes of uses of the devices.

\begin{acknowledgments}
We are indebted to Tai Tsun Wu, whose presence pervades these pages,
and to Howard E. Brandt for continuing discussions concerning quantum
mechanics.  We thank Dionisios Margetis for willing and
effective analytic help.

This work was supported in part by the Air Force Research Laboratory and 
DARPA under Contract F30602-01-C-0170 with BBN Technologies.
\end{acknowledgments}

\newpage
\appendix

\section{Solution to Model $\alpha$ of a 1-bit recording device}

Starting with Eq.\ (\ref{eq:pairphys}), and writing $t$ as
the time parameter times a dimensionless ``$t$'' and $x$ and $y$ as
the distance parameter times dimensionless ``$x$'' and ``$y$,''
respectively, we obtain
\begin{equation}i\,\frac{\partial \;}{\partial t}\,\psi(x,y,t) =
\frac{1}{2}
\left(-\frac{\partial^2 \;}{\partial x^2} -\frac{\partial^2
\;}{\partial y^2} - x^2 - y^2 + \frac{\lambda}{2}
(x-y)^2\right)\psi(x,y,t).  \label{eq:pair}
\end{equation} 
This equation is solved by introducing a non-local coordinate change:
\begin{equation} u = \frac{x +y}{\sqrt{2}} 
\quad\text{and}\quad v = \frac{x-y}{\sqrt{2}}.
\end{equation}
With this change of variable, Eq.\ (\ref{eq:pair}) becomes
\begin{equation}i\,\frac{\partial \;}{\partial t}\,\psi(u,v,t) =
\frac{1}{2}\left(-\frac{\partial^2 \;}{\partial u^2} -\frac{\partial^2
\;}{\partial v^2} - u^2 + (\lambda-1) v^2\right)\psi(u,v,t),
\label{eq:uvpair}
\end{equation} for which separation of variables is immediate,
so the general solution is a sum of products, each of the form
\begin{equation}
\psi(u,v,t) = \phi(u,t)\chi(v,t). 
\label{eq:factor}
\end{equation}
The function $\phi$ satisfies its own Schr\"odinger  
equation,  
\begin{equation}
\left(i\,\frac{\partial \;}{\partial t} + \frac{1}{2}\,\frac{\partial^2
\;}{\partial u^2} + \frac{u^2}{2}\right)\phi(u,t) = 0,  
\label{eq:phieq}
\end{equation}
which is the quantum-mechanical equation for an unstable harmonic
oscillator, while $\chi$ satisfies
\begin{equation}
\left(i\,\frac{\partial \;}{\partial t} + \frac{1}{2}\,
\frac{\partial^2 \;}{\partial v^2} + \frac{1- \lambda}{2}\,
v^2\right)\chi(v,t)  = 0 , \label{eq:chieq}
\end{equation}
which varies in its interpretation according to the value of
$\lambda$, as follows: (a) for $\lambda < 1$ it expresses an unstable
harmonic oscillator; (b) for $\lambda = 1$ it expresses a free particle,
and (c) for $\lambda > 1$ it expresses a stable harmonic oscillator.  The
last case will be of interest when we compare behavior of the model
with an experimental record.

By translating Eq.\ (\ref{eq:initxy})\ into $(u,v)$-coordinates, one
obtains initial conditions
\begin{eqnarray}
\phi(u,0) &\mcl=\mcl&
\pi^{-1/4}b^{-1/2}\exp\left\{-\frac{(u-\sqrt{2}c)^2}{2b^2}\right\}, \\
\chi(v,0) &\mcl=\mcl& \pi^{-1/4}b^{-1/2}\exp\left\{-\frac{v^2}{2b^2}
\right\}.
\end{eqnarray}
The solution to Eq.\ (\ref{eq:phieq}) with these
initial conditions is given by Barton \cite{barton}; we deal with
$\phi$ and $\chi$ in order.  From (5.3) of Ref.~\cite{barton}, one has
\begin{equation}
|\phi(u,t)|^2 = \frac{1}{\pi^{1/2}B_1(t)}\exp\left\{\frac{-(u -
c\sqrt{2}\cosh
 t)^2}{B_1^2(t)}\right\},
\label{eq:phiprob}
\end{equation}
where 
\begin{equation}
B_1^2(t) = b^2 \left[1 + \left(\frac{1}{b^4} +1\right)\sinh^2 t\right].
\label{eq:Bdef}
\end{equation}
Similarly, integrating the Green's function for the stable oscillator
($\lambda > 1)$ over the initial condition for $\chi$ yields
\begin{equation}
|\chi(v,t)|^2 =
\frac{1}{\pi^{1/2}B_2(t)}\exp\left\{\frac{-v^2}{B_2^2(t)}\right\},
\end{equation}
where 
\begin{equation}
B_2^2(t) = b^2\left[1 + \left(\frac{1}{b^4(\lambda-1)}-1\right)
  \sin^2\sqrt{\lambda-1}\,t\right]
\label{eq:B2def}
\end{equation}
Multiplying these and changing back to $(x,y)$-coordinates yield the
joint probability density 
\begin{equation}
|\psi(x,y,t)|^2 = \frac{1}{\pi B_1(t)B_2(t)}\,
\exp\left\{-\left(\frac{x+y}{\sqrt{2}} - c\sqrt{2}\cosh
  t\right)^2\Bigg/ B^2(t)- \frac{(x-y)^2}{2B^2_2(t)}\right\}.
\label{eq:jointxy}
\end{equation}
The probability of two detections disagreeing is the integral of this
density over the second and fourth quadrants of the $(x,y)$-plane.
This is most conveniently carried out in $(u,v)$-coordinates.  For the
especially interesting case of $c =0$ (and $\lambda > 1)$, this
integral can be transformed into
\begin{eqnarray}
\Pr(F1\text{ and }F2\text{ disagree at }t) &\mcl=\mcl& \frac{1}{\pi
B_1(t) B_2(t)}\int^{\infty}_{-\infty} dv \int^v_{-v}du\,
\exp\left\{\frac{-u^2}{B_1^2(t)} - \frac{v^2}{B_2^2(t)}\right\} \nonumber
\\ &\mcl=\mcl&
\frac{4}{\pi}\int_0^{\infty}e^{-V^2}\, dV\int_0^{B_2(t) V/B_1(t)}dU\,
e^{-U^2} \nonumber \\ &\mcl=\mcl&
\frac{2}{\pi}\tan^{-1}\left(\frac{B_2(t)}{B_1(t)}\right)
\nonumber \\ &\mcl=\mcl& \frac{2}{\pi}\tan^{-1}\left( \frac{\displaystyle
1 + \left[\frac{1}{b^4(\lambda-1)}-1\right]\sin^2\sqrt{\lambda-1}\,
t}{\displaystyle 1 +\left(\frac{1}{b^4}+1\right)\sinh^2
t}\right)^{1/2}.\qquad
\label{eq:edge1}
\end{eqnarray}
It is easy to check that this formula works not only when $\lambda >
1$ but also for the case $\lambda < 1$.  For $0 < \lambda <1$,
the numerator takes on the same form as the denominator, but with
a slower growth with time, so that the probability of disagreement
still decreases with time exponentially, but more slowly.

Converting Eq.\ (\ref{eq:jointxy}) from dimensionless back to physical
time and distance variables results in Eq.\ (\ref{eq:jointxyphys}),
and similarly Eq.\ (\ref{eq:edge1}) leads to Eq.\ (\ref{eq:edge1phys}).

\eject
\begin{figure}
\centerline{Figure Captions}
\vspace{35pt}
\caption{Flip-flop exposed to race between signal going high and clock going low.
}
\vspace{25pt}
\caption{Flip-flops $F1$ and $F2$ used to read $F0$ after a delay $t$.}
\vspace{25pt}
\caption{Probability of disagreement vs.\ settling time.}
\end{figure}
\newpage
\clearpage
\setcounter{figure}{0}
\begin{figure}[t]
\begin{center}
\vspace*{135pt}
   \begin{tabular}{c}
   \includegraphics[width=5.75in]{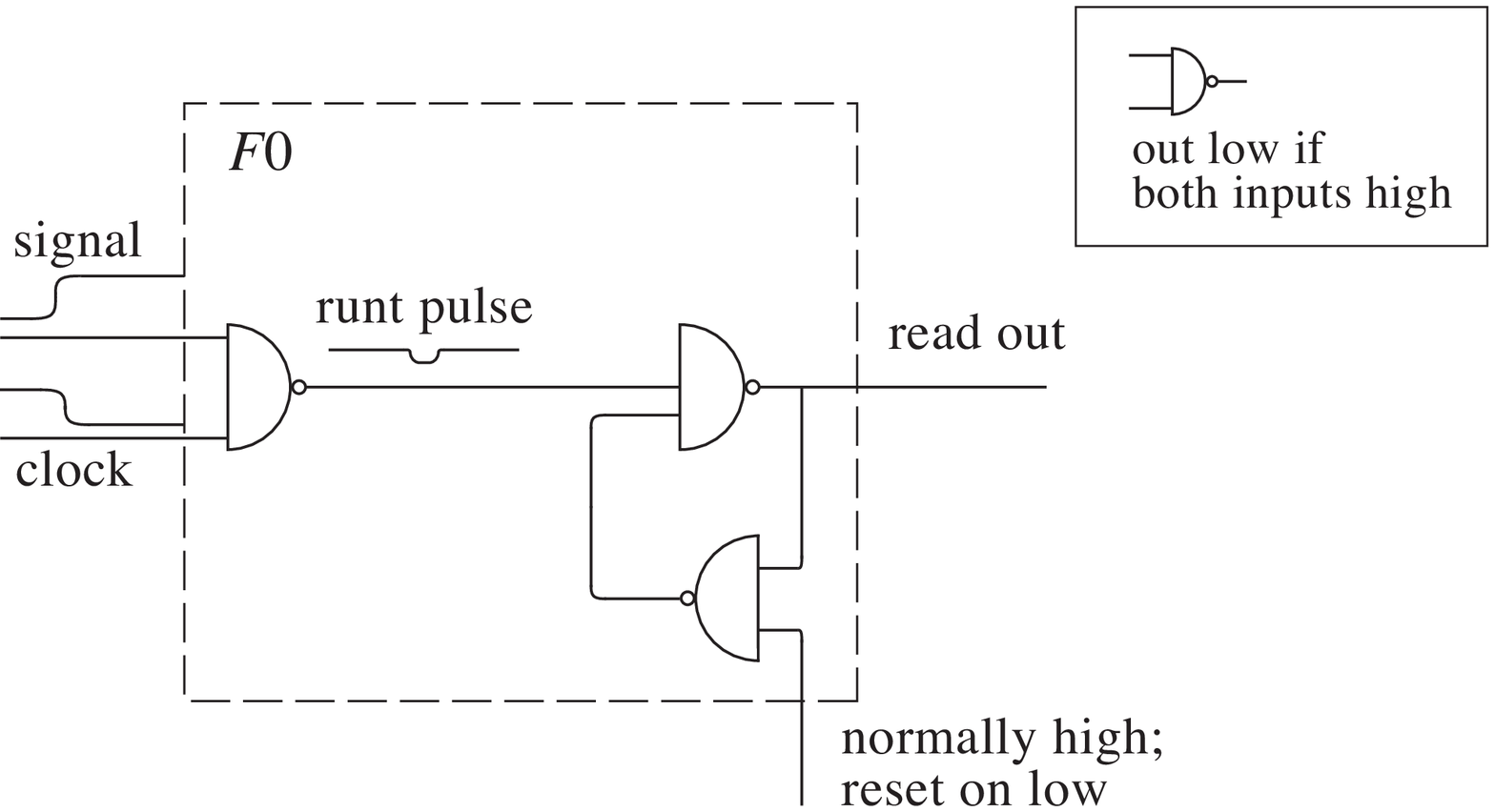}
   \end{tabular}
   \end{center}
\caption[]   { \label{fig:1} 
Flip-flop exposed to race between signal going high and clock going low.
}
   \end{figure} 
\newpage
\clearpage
\eject
\begin{figure}[t]
\begin{center}
\vspace*{120pt}
   \begin{tabular}{c}
   \includegraphics[width=5.75in]{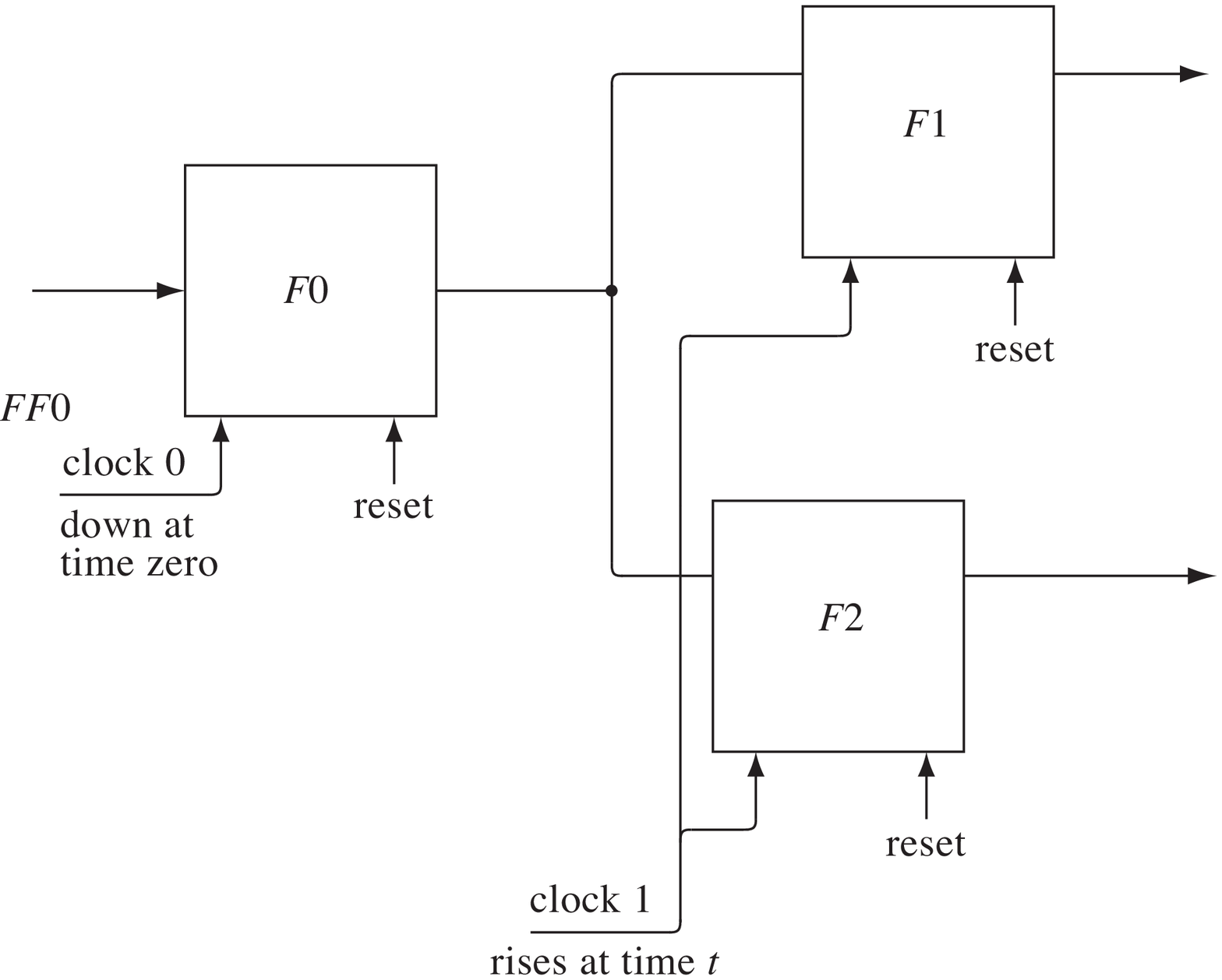}  
   \end{tabular}
   \end{center}
\caption[]   { \label{fig:2} 
Flip-flops $F1$ and $F2$ used to read $F0$ after a delay $t$.
}
   \end{figure} 
\newpage
\clearpage
\eject
\begin{figure}[t]
\begin{center}
\vspace*{100pt}
   \begin{tabular}{c}
   \includegraphics[width=5.75in]{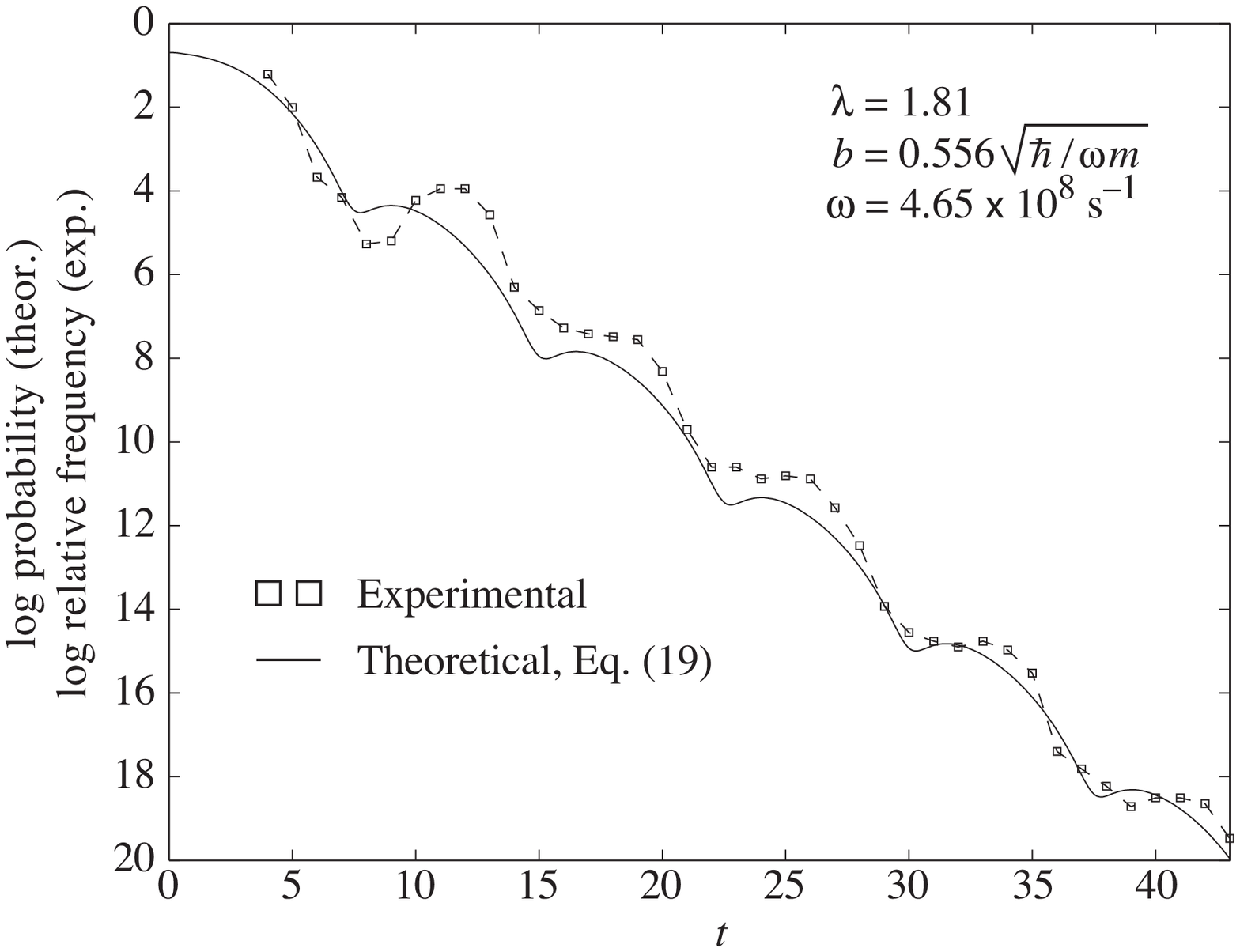}
   \end{tabular}
   \end{center}
\caption[]   { \label{fig:3} 
Probability of disagreement vs.\ settling time.
}
   \end{figure} 

\end{document}